\documentclass[final,3p,times,twocolumn]{elsarticle}

\usepackage{lineno,hyperref,amsmath,amssymb,color}
\hypersetup{colorlinks,bookmarksopen,bookmarksnumbered,citecolor=rossos,
linkcolor=black,pdfstartview=FitH,urlcolor=blus}
\modulolinenumbers[5]
\biboptions{numbers,sort&compress}

\journal{the arXiv}

\begin{document}

\begin{frontmatter}

\title{Axion cosmology, lattice QCD and the dilute instanton gas}

\author[wuppertal]{Sz.~Borsanyi}

\author[desy]{M.~Dierigl}

\author[wuppertal,juelich,budapesta]{Z.~Fodor}

\author[budapesta,budapestb]{S.D.~Katz}

\author[regensburg,juelich]{S.W.~Mages}

\author[budapesta,budapestb,kavli]{D.~Nogradi}

\author[zaragoza,mpi]{J.~Redondo}

\author[desy]{A.~Ringwald}

\author[wuppertal,juelich]{K.K.~Szabo}

\address[wuppertal]{Department of Physics, Wuppertal
    University, Gaussstrasse 20, D-42119 Wuppertal, Germany}

\address[desy]{Deutsches Elektronen-Synchrotron DESY, Notkestrasse 85, D-22607 Hamburg, Germany}

\address[juelich]{IAS/JSC, Forschungszentrum J\"ulich, D-52425
    J\"ulich, Germany}

\address[budapesta]{Institute for Theoretical Physics, E\"otv\"os
    University, P\'azm\'any Peter s\'etany 1/A, H-1117 Budapest, Hungary}

\address[budapestb]{MTA-ELTE Lend\"ulet Lattice Gauge Theory Research Group, Budapest, Hungary}

\address[regensburg]{University of Regensburg,  D-93053 Regensburg, Germany}

\address[kavli]{Kavli Institute for Theoretical Physics, University of California Santa Barbara, CA 93106-4030, USA}

\address[zaragoza]{Departamento de F\'isica Te\'orica, Universidad de Zaragoza, Pedro Cerbuna 12, E-50009, Zaragoza, Espa\~na}

\address[mpi]{Max-Planck-Institut f\"ur Physik (Werner-Heisenberg-Institut),
F\"ohringer Ring 6, D-80805 M\"unchen, Germany 
\\[1.5ex] {\it {\small Report number: DESY 15-151}}}

\begin{abstract}
Axions are one of the most attractive dark matter candidates. The evolution of their number density in the early universe can be determined by calculating the topological susceptibility $\chi(T)$ of QCD as a function of the temperature. Lattice QCD provides an {\it ab initio} technique to carry out such a calculation. A full result needs two ingredients: physical quark masses and a controlled continuum extrapolation from non-vanishing to zero lattice spacings. We determine $\chi(T)$ in the quenched framework (infinitely large quark masses) and extrapolate its values to the continuum limit. The results are compared with the prediction of the dilute instanton gas approximation (DIGA). A nice agreement is found for the temperature dependence, 
whereas the overall normalization of the DIGA result still differs from the non-perturbative continuum extrapolated lattice results by a factor of order ten. We discuss the consequences of our findings for the prediction of the amount of axion dark matter. 
\end{abstract}


\end{frontmatter}

\section{Introduction}

One of the greatest puzzles in particle physics is the nature of dark matter. A prominent particle candidate for the 
latter is the 
axion $A$ \cite{Weinberg:1977ma,Wilczek:1977pj}:  a pseudo Nambu-Goldstone boson arising from the breaking of a hypothetical global chiral 
$U(1)$ extension \cite{Peccei:1977hh} of the Standard Model  at an energy scale $f_A$ much larger than the electroweak scale. 

A key input for the prediction of the amount of axion dark matter \cite{Preskill:1982cy,Abbott:1982af,Dine:1982ah} is its potential as a function of the temperature, $V(A,T)$. 
It is related to the free energy density in QCD, 
$F(\theta ,T)\equiv - \ln Z (\theta,T)/{\mathcal V}$, via 
\begin{eqnarray}
V(A,T) \equiv - \frac{1}{\mathcal V} \ln \left[\frac{Z (\theta,T)}{Z (0,T)}\right]|_{\theta = A/f_A}
,
\label{axion_potential_free_energy_density}
\end{eqnarray} 
where $\mathcal{V}$ is the Euclidean space-time volume. The angle $\theta$ enters the Euclidean QCD Lagrangian via the additional term involving the topological charge density $q(x)$, 
\begin{equation}
-i \theta q (x) \equiv -i \theta 
\frac{\alpha_s}{16\pi}   \epsilon_{\mu\nu\rho\sigma} F_{\mu\nu}^a (x) F^a_{\rho\sigma} (x) ,
\end{equation}
with $F_{\mu\nu}^a$ being the gluonic field strength and $\alpha_s\equiv g_s^2/(4\pi)$ the fine structure constant of strong interactions.

On general grounds, the free energy density and thus the axion potential has 
an absolute minimum at $\theta = A/f_A = 0$. In fact, this is the reason why in this extension of the Standard Model there is no strong CP problem \cite{Peccei:1977hh}. The curvature around this minimum determines the axion mass 
$m_A$ at finite temperature,
\begin{equation}
m_A^2(T) \equiv \frac{\partial^2 V(A,T)}{\partial A^2}|_{A=0}   = \frac{\chi (T)}{f_A^2} ,
\end{equation}
in terms of the topological susceptibility, i.e.  
the variance of the $\theta =0$ topological charge distribution,
\begin{equation}
\chi (T) \equiv \int d^4x \langle q (x) q (0)\rangle_T|_{\theta =0}
= \lim_{{\mathcal V}\to \infty}  \frac{\langle Q^2\rangle_T|_{\theta =0}}{\mathcal V}  
,
\end{equation}
where $Q\equiv\int d^4x q(x)$.
Similarly, self-interaction terms in the potential, e.g. the $A^4$ term occurring in the expansion of the free energy density
around $\theta = A/f_A =0$, 
\begin{equation}
\label{axion_potential_density_exp}
V(A,T)  =  \frac{1}{2} \chi (T)\theta^2 \left[ 1 + b_2(T) \theta^2 + \ldots \right]|_{\theta = A/f_A} , 
\end{equation}
are determined by higher moments,
\begin{eqnarray}
b_2 (T) = -\, \frac{ \langle Q^4 \rangle_T - 3  \langle Q^2 \rangle^2_T}{ 
12 \langle Q^2 \rangle_T }\mid_{\theta =0} .
\end{eqnarray}
These non-perturbative quantities enter in a prediction of the lower bound on the fractional contribution of axions to the observed cold dark matter as follows\footnote{
This is a rewriting of equation (2.10) of Ref. \cite{Arias:2012az} where we have used $\chi(0)=(m_A f_A)^2 \simeq 3.6\times 10^{-5}$ GeV$^{4}$ from the chiral Lagrangian to express $f_A$ in terms of $m_A$, the zero temperature mass, that is itself a function of $T_{\rm osc}$ and the ratio $\chi(T)/\chi(0)=(m_A(T)/m_A)^2$ through the condition for the onset of the oscillations, $m_A(T_{\rm osc})=3 H(T_{\rm osc})$ with $H^2=8\pi^3 G_N g_{*R}(T) T^4/90$ the Hubble expansion rate.} 
\begin{equation}
R_A \gtrsim 10^5 \langle {\theta^2}\rangle \, f(\langle {\theta^2}\rangle )\frac{\rm GeV^3}{g_{*S}(T_{\rm osc})T_{\rm osc}^3}\sqrt{\frac{\chi(T_{\rm osc})}{\chi(0)}}, 
\label{misalignment}
\end{equation} 
where $\langle \theta^2\rangle$ is the variance of the spatial distribution of the initial values of the axion field $A/f_A$ 
before the formation of the dark matter condensate by the misalignment mechanism, which occurs, 
when the Hubble expansion rate gets of the order of the
axion mass, $m_A(T_{\rm osc})=3 H(T_{\rm osc})$, i.e.\  at a temperature 
\begin{equation}
T_{\rm osc} \simeq 
\frac{50\ {\rm GeV}}{g^{1/4}_{*R}(T_{\rm osc})} 
\left(\frac{m_A}{{\rm \mu eV}}\right)^{1/2} \left( \frac{\chi(T_{\rm osc})}{\chi(0)}\right)^{1/4}
.
\end{equation} 
The function $f(\langle {\theta^2}\rangle )$ is taking into account anharmonicity effects
arising from the self-interaction terms in $V(A,T)$ and depends on the specific form of the potential.  The functions 
$g_{*R}(T)$ and $g_{*S}(T)$ denote the effective number of relativistic energy and entropy degrees of freedom, respectively. 

What is urgently needed for axion cosmology is thus a precise determination of the topological susceptibility and higher moments of the topological charge distribution. In this context, most predictions have been entirely based on the semi-classical 
expansion of the Euclidean path integral of finite temperature QCD around a dilute gas of instantons -- finite action minima of the Euclidean action with unit topological charge -- see e.g. 
Ref. \cite{Turner:1985si} for an early exhausting study and Ref. \cite{Bae:2008ue} 
for a recent update concerning the quark masses.  A comparative study of these predictions based on the dilute instanton gas approximation (DIGA) has been carried out in Ref. \cite{Wantz:2009it}, where also an analysis in terms of a phenomenological instanton liquid model (IILM) \cite{Wantz:2009mi} is presented. 
However, up to now, in all DIGA investigations of the topological susceptibility only the one-loop expression in the 
expansion around the instanton background field was used. This results in a strong renormalization scale dependence 
and thus large uncertainties which were neglected in the previous DIGA based predictions. 
In fact, in the temperature range $T_{\rm osc}\sim {\rm GeV}$ of interest, one expects a large uncertainty in the overall normalization due to the neglection of higher order loop effects, since in this region $\alpha_s(T_{\rm osc})$ is not small.
We will exploit in this letter both the one-loop DIGA result as well as its two-loop renormalization group improved (RGI) version in order to study the theoretical uncertainties arising from higher loop corrections. Most importantly, we compare these
predictions with the outcome of our lattice based fully non-perturbative results.

Actually, there have been a number of 
lattice calculations of $\chi (T)$ and $b_2(T)$ at temperatures below or slightly above the QCD phase transition, mostly in quenched QCD, see e.g. \cite{Bonati:2013tt,Berkowitz:2015aua,Kitano:2015fla}.
Here we go beyond those lattice calculations. We extend
the available temperature range and carry out a controlled continuum extrapolation for this extended range. In addition, note that there has been no quantitative investigation whether and where the lattice results turn into the DIGA results.  
We will present our new high-quality lattice data for the topological susceptibility in 
quenched QCD (i.e. neglecting the effects of light quarks) and compare them quantitatively to the DIGA result.

\section{Axion potential coefficients from the lattice}

On the lattice, the topological susceptibility is measured on the torus as the second moment of the distribution of the global topological charge 
\[ \chi_t=\langle Q^2 \rangle/{\mathcal V},\] 
where $Q$ is any
renormalized discretization of the global topological charge, and $\mathcal V$ is the
four-volume of the lattice.  There are a lot of different fermionic and gluonic
definitions of $Q$ available. We choose a gluonic definition based on the
Wilson flow \cite{Luscher:2010iy,Luscher:2011bx}, which has the correct continuum limit similarly to the fermionic definitions but is numerically a lot cheaper. In particular we evolve our gauge field configurations with the Wilson plaquette action to a flow time $t$ and define the global topological
charge as the integral over the clover definition of the topological charge
density. This definition gives a properly renormalized observable when the
flow time $t$ is fixed in physical units.

\begin{figure}
\includegraphics[width=\columnwidth]{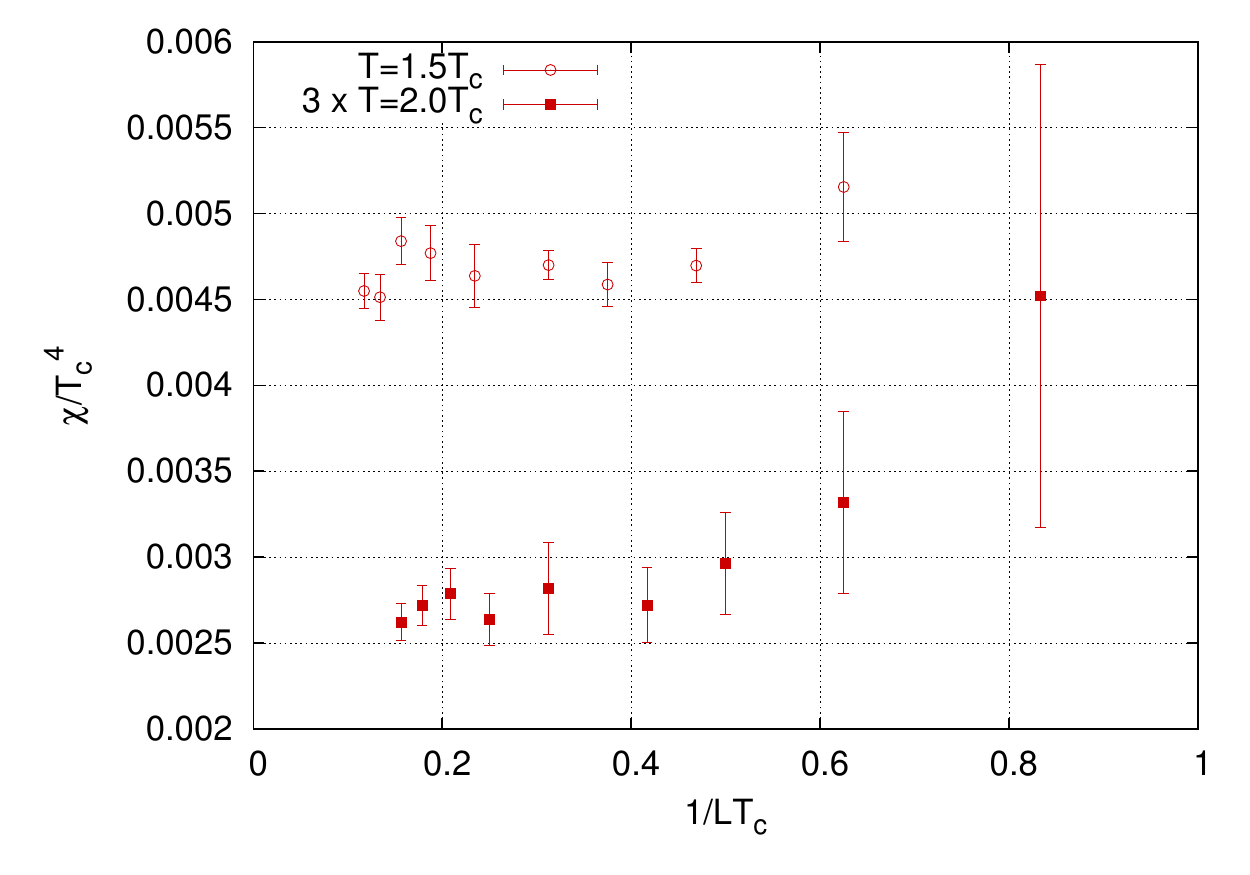}
\caption{Demonstration that volume is
sufficiently large to have negligible finite volume corrections on $Q^2$. The data for $T/T_c=2$ is multiplied by 3 for better visibility of the comparison.}
\label{fi:finv}
\end{figure}

We use a tree-level Symanzik improved gauge action.  Our temperatures range
from below $T_c$ up to $4T_c$. Here $T_c$ is the critical temperature, which is
the quantity used for scale setting. The critical temperatures for different
lattice spacings were determined in earlier work \cite{Cella:1994sx,Borsanyi:2012ve}.  
For the whole temperature
range we keep the spatial lattice size approximately at $L=2/T_c$. We checked
with dedicated high volume runs at $1.5T_c$ and $2T_c$ that this volume is
sufficiently large to have negligible finite volume corrections on $Q^2$. This
is shown in Fig.\ref{fi:finv}.  Our spatial geometry
is $L\times L\times 2L$ to enable tests of subvolume methods which will be
reported separately -- here we only use the full volume.  For all temperatures
we have three lattice spacings $(aT)^{-1}=5,6,8$ to be able to perform an
independent continuum extrapolation for every temperature. The local
heatbath/overrelaxation algorithm is used for the update, one sweep consists of
1 heatbath and 4 overrelaxation steps.  We found that the autocorrelation time
of the topological charge depends weakly on $(aT)$, i.e.\ if the temperature is
increased by decreasing the lattice spacing.  The number of update sweeps
between measurements was chosen in accordance with the autocorrelation time.
Tab.\ \ref{ta:sim} lists the simulation points with the number of sweeps.

\begin{table}
    \centering
    \begin{tabular}{|c|c|c|c|}
	\hline
	$T/T_c$ & $N_t$ & $N_s$ & $N_{sweeps}$ \\\hline
	\hline
0.9 & 5 & 12 & 32K \\
& 6 & 12 & 48K \\
& 8 & 16 & 170K \\
\hline
1.0 & 5 & 12 & 48K \\
& 6 & 12 & 64K \\
& 8 & 16 & 180K \\
\hline
1.1 & 5 & 12 & 48K \\
& 6 & 16 & 160K \\
& 8 & 20 & 330K \\
\hline
1.3 & 5 & 16 & 64K \\
& 6 & 16 & 220K \\
& 8 & 24 & 550K \\
\hline
1.5 & 5 & 16 & 96K \\
& 6 & 20 & 210K \\
& 8 & 24 & 660K \\
\hline
1.7 & 5 & 20 & 260K \\
& 6 & 20 & 300K \\
& 8 & 28 & 700K \\
\hline
2.0 & 5 & 20 & 280K \\
& 6 & 24 & 510K \\
& 8 & 32 & 840K \\
\hline
2.3 & 5 & 24 & 420K \\
& 6 & 28 & 530K \\
& 8 & 36 & 900K \\
\hline
2.6 & 5 & 28 & 710K \\
& 6 & 32 & 2600K \\
& 8 & 44 & 5000K \\
\hline
3.0 & 5 & 32 & 810K \\
& 6 & 36 & 7400K \\
& 8 & 48 & 1700K \\
\hline
3.5 & 5 & 36 & 910K \\
& 6 & 44 & 2100K \\
& 8 & 56 & 5100K \\
\hline
4.0 & 5 & 40 & 810K \\
& 6 & 48 & 2200K \\
& 8 & 64 & 4900K \\
	\hline
    \end{tabular}
    \caption{\label{ta:sim} List of simulation points: temperatures, lattice sizes $N_t=1/(aT)$, $N_s=L/a$, and number of sweeps are given.}
\end{table}

\begin{figure}
\includegraphics[width=\columnwidth]{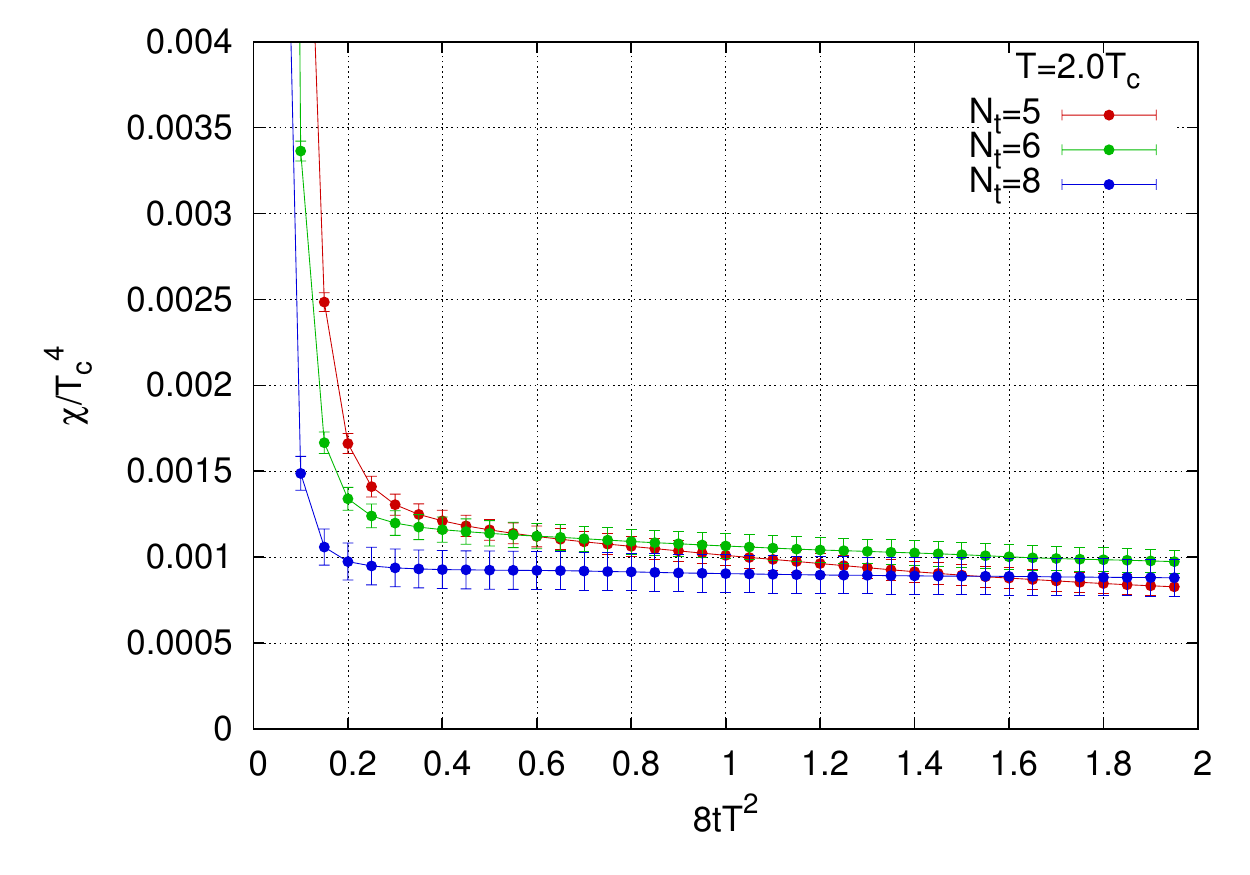}
\caption{Demonstration that Wilson flow/lattice renormalization are under control.}
\label{fi:flow_plot}
\end{figure}

We integrated the Wilson flow numerically to a maximum flow-time of 
about $8t\approx 1/(2 T_c^2)$ for all temperatures.  Figure \ref{fi:flow_plot}
gives the dependence of the susceptibility on the flow time for
$T=2T_c$. While in the continuum limit the result is independent of the choice
of the flow time $t$, different choices have very different lattice artefacts.
For small flow times the different lattice spacings give very different
results.  For larger flow times the expected plateau behavior can be observed
for each lattice spacing and the lattice artefacts also decrease significantly.
The choice of the flow time brings in some arbitrariness into the analysis, 
however the continuum result should not depend on this choice once $t$ is fixed in physical units.  But
this is certainly a subleading source of error compared to the statistical
error due to the rare topology tunneling events at high temperatures.  In this
analysis we choose a temperature dependent flow time for the evaluation of
$Q^2$ as
\begin{align}
	8t=\begin{cases}
	1/ (1.5T_c)^2, & T<1.5\\
	1/T^2, & T\geq1.5\end{cases}.
\end{align}
For low/high temperatures this means a temperature independent/dependent flow
time. This choice is safely in the expected plateau region for all
temperatures.

\begin{figure}[h]
\includegraphics[width=\columnwidth]{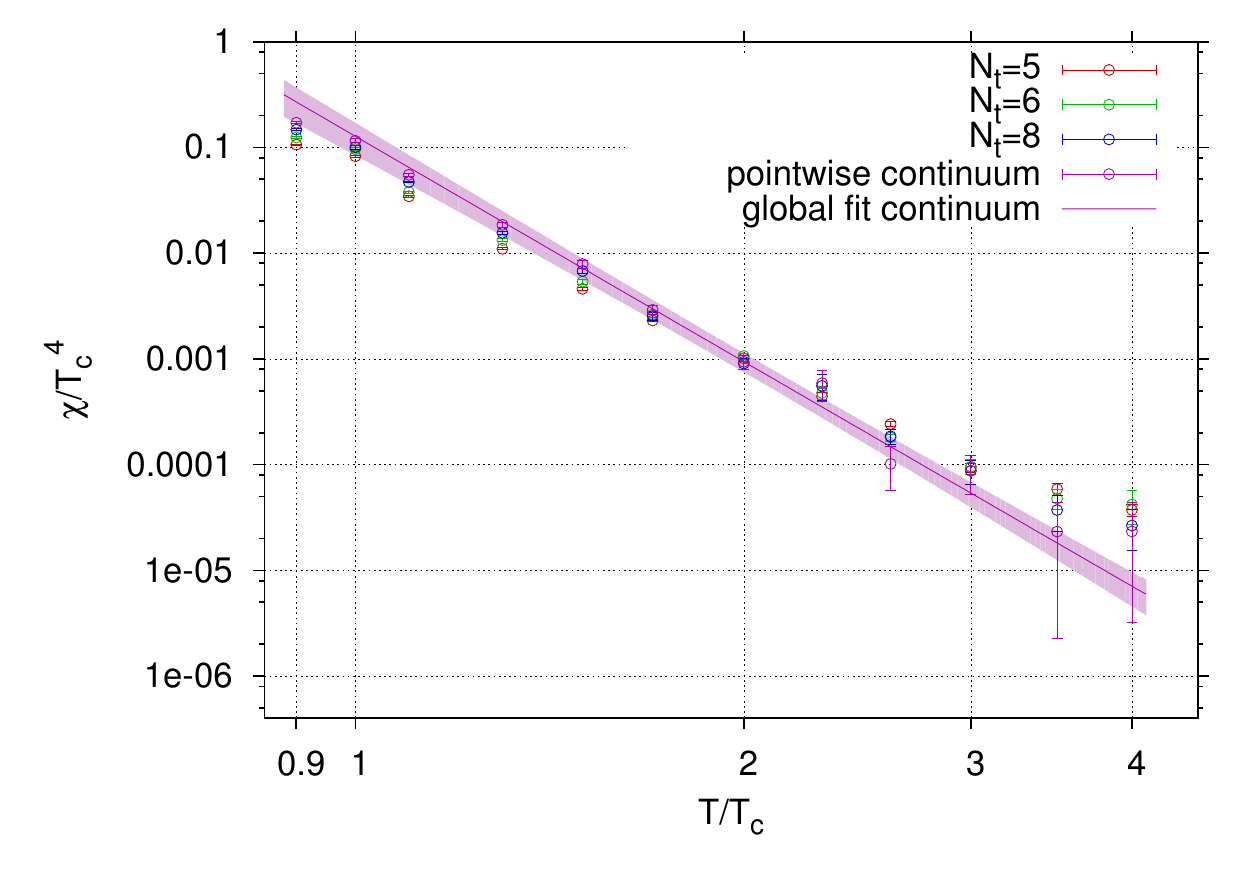}
\caption{Lattice data on the topological susceptibility at $N_t=5,6,8$ and lattice continuum extrapolation together with fit of simple power law
\label{fi:bare_plot}}
\end{figure}

The resulting values for the susceptibility are plotted in Fig.
\ref{fi:bare_plot}. This plot also gives the result of a global continuum
extrapolation using a set of temperatures and the 6-parameter power law ansatz
\begin{align}
    \chi_t=(\chi_0+\chi_0' a^2 )\left(\frac{T}{T_0+T_0' a^2}\right)^{b + b' a^2},
\end{align}
where $\chi_0$, $T_0$, and $b$ are fit parameters giving the continuum limit.
$\chi_0'$, $T_0'$, and $b'$ are fit parameters describing the deviation from the
continuum limit. The fit parameter $T_0$ is included as a consistency check and
should give 1 in units of $T_c$. This is satisfied by the fit result.
The variation between different choices for the starting temperature
of the fit range $T_{min}/T_c=1.3,1.5,1.7$ gives an estimate of
the systematic error of the result. The best fit parameters  are 
\begin{align}
    \chi_0= 0.11(2)(1), \quad b= -7.1(4)(2), \quad T_0= 1.02(5)(2),
\label{lattice_fitting_parameters}
\end{align}
where the first error is the statistical, the second is the systematic.
The point-wise continuum extrapolation is consistent with the global fit, evidently the
latter has smaller errors for large temperatures.
Note that though a controlled continuum extrapolation is possible using three
lattice spacings, estimating the systematic uncertainty of this extrapolation
would require at least one more lattice resolution.

In a recent analysis \cite{Berkowitz:2015aua} the topological susceptibility was
calculated using the techniques \cite{DelDebbio:2002xa,Durr:2006ky}. The
calculation was carried out at two temporal extensions, corresponding to two
lattice spacings at each temperature. The exponent $b = -5.64(4)$ was found
which differs from our value. Note however, that our temperature range is
larger, thus we are closer to the applicability range of the DIGA. Furthermore,
the two lattice spacings were not sufficient for a controlled continuum
extrapolation thus uncertainties related to this final step are not included
in the result of \cite{Berkowitz:2015aua}. 

\begin{figure}
\includegraphics[width=\columnwidth]{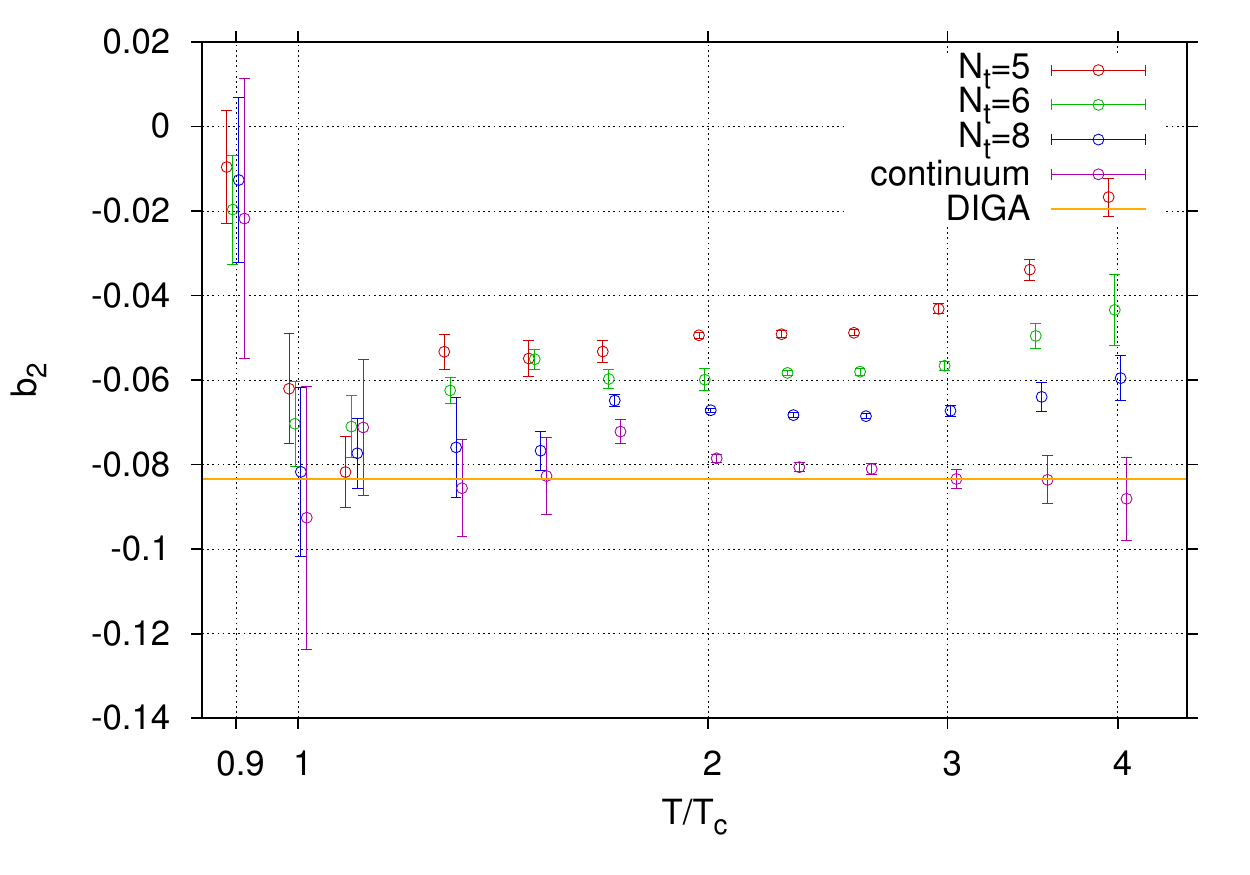}
\caption{Lattice data on the anharmonicity coefficient $b_2$ of the axion potential compared to its DIGA prediction.
The data points are shifted a bit horizontally for better visibility.
\label{fi:b2_plot}}
\end{figure}

We have also determined the second important coefficient $b_2$ of the axion potential, 
characterizing its anharmonicity, 
by measuring the observable 
\[ b_{2\,,t} = -\, \frac{ \langle Q^4 \rangle - 3  \langle Q^2 \rangle^2}{ 
12 \langle Q^2 \rangle }.\] 
The result is plotted in Fig. \ref{fi:b2_plot}. 

\section{Comparison between lattice and DIGA results}

In this section, we confront the lattice results with the ones obtained from the DIGA framework.  
For the sake of completeness let us collect first the available formulas for the latter  \cite{Hooft:1976fv,Callan:1977gz,Bernard:1979qt,Pisarski:1980md,Gross:1980br,Luscher:1981zf,Morris:1984zi,Ringwald:1999ze}. 
At very high temperatures, far above the QCD phase transition, it makes sense to infer the $\theta$ dependence of QCD from the DIGA, in which the partition function can be written as \cite{Callan:1977gz}, 
\begin{equation}
 Z (\theta ,T) \simeq \sum_{n_I,n_{\bar I}}\frac{1}{n_I!n_{\bar I}!} Z_I^{n_I+n_{\bar I}} (T)\exp\left[i\theta (n_I - n_{\bar I})\right]\,,
\end{equation}
where $Z_I=Z_{\bar I}$ is the contribution arising from the expansion of the path integral around a single instanton $I$ (anti-instanton ${\bar I}$). It follows directly that the potential has the form
\begin{equation}
\label{free_energy_density_digq}
V(A,T)    \simeq  \chi (T)\left( 1 - \cos\theta \right)|_{\theta=A/f_A},
\end{equation}
from which one infers 
\begin{equation}
b_2(T) \simeq - \frac{1}{12}.
\end{equation} 
This can be confronted right-away  with our lattice results, cf.\ Fig.\ \ref{fi:b2_plot}. 
Similar to Ref.\ \cite{Bonati:2013tt} we find that the prediction from the DIGA for $b_2$ is reached already at surprisingly low 
values of $T/T_c\gtrsim 1$. 

The whole temperature dependence of the axion potential arises in the DIGA 
through the topological susceptibility, which  in this case is explicitly given by  
\begin{equation}
\chi (T) \simeq 
\frac{Z_I(T)+Z_{\bar I}(T)}{\mathcal V} =  2 \int_0^\infty d\rho\, D(\rho )\,G(\pi\rho T ),
\label{eq:top_susc_diga}
\end{equation}
in terms of the instanton size distribution at zero temperature, 
$D(\rho )$, and a factor $G(\pi\rho T)$ taking into account finite temperature effects. The former is known in the framework of the semiclassical expansion around the instanton for small 
$\alpha_s(\mu_r)\ln(\rho\,\mu_r)$ and $\rho\,m_i(\mu_r)$, where $\alpha_s$ is the strong coupling, $\mu_r$ is the renormalization
scale and $m_i(\mu_r)$ are the running quark masses. 
To one-loop accuracy, it is given 
by\footnote{For quenched QCD the number of light quarks is $n_f=0$; the general formula can be found explicitly in e.g.\
Ref. \cite{Ringwald:1999ze}, 
which contains a pioneering confrontation of cooled lattice data on $D(\rho )$ with the two-loop RG improved
DIGA result.
}
\begin{eqnarray}
\label{dens}
D({\rho})&\equiv & 
\frac{d_{\overline{\rm MS}}}{\rho^5}
\left(\frac{2\pi}{\alpha_{\overline{\rm MS}}(\mu_r)}\right)^{6} 
\exp{\left(-\frac{2\pi}{\alpha_{\overline{\rm MS}}(\mu_r)}\right)}
\\[1.5ex] \nonumber
&& \times  
(\rho\,
\mu_r)^{\beta_0} 
\left[ 1 +\mathcal{O}(\alpha_{\overline{\rm MS}}(\mu_r)) \right]
,
\end{eqnarray}
with 
\begin{eqnarray}
 d_{\overline{\rm MS}}=\frac{{\rm e}^{5/6}}{\pi^2}\,{\rm
 e}^{-4.534122}; \hspace{3ex}
\beta_0=11 .
\end{eqnarray}

\begin{figure}[t]
\includegraphics[width=\columnwidth]{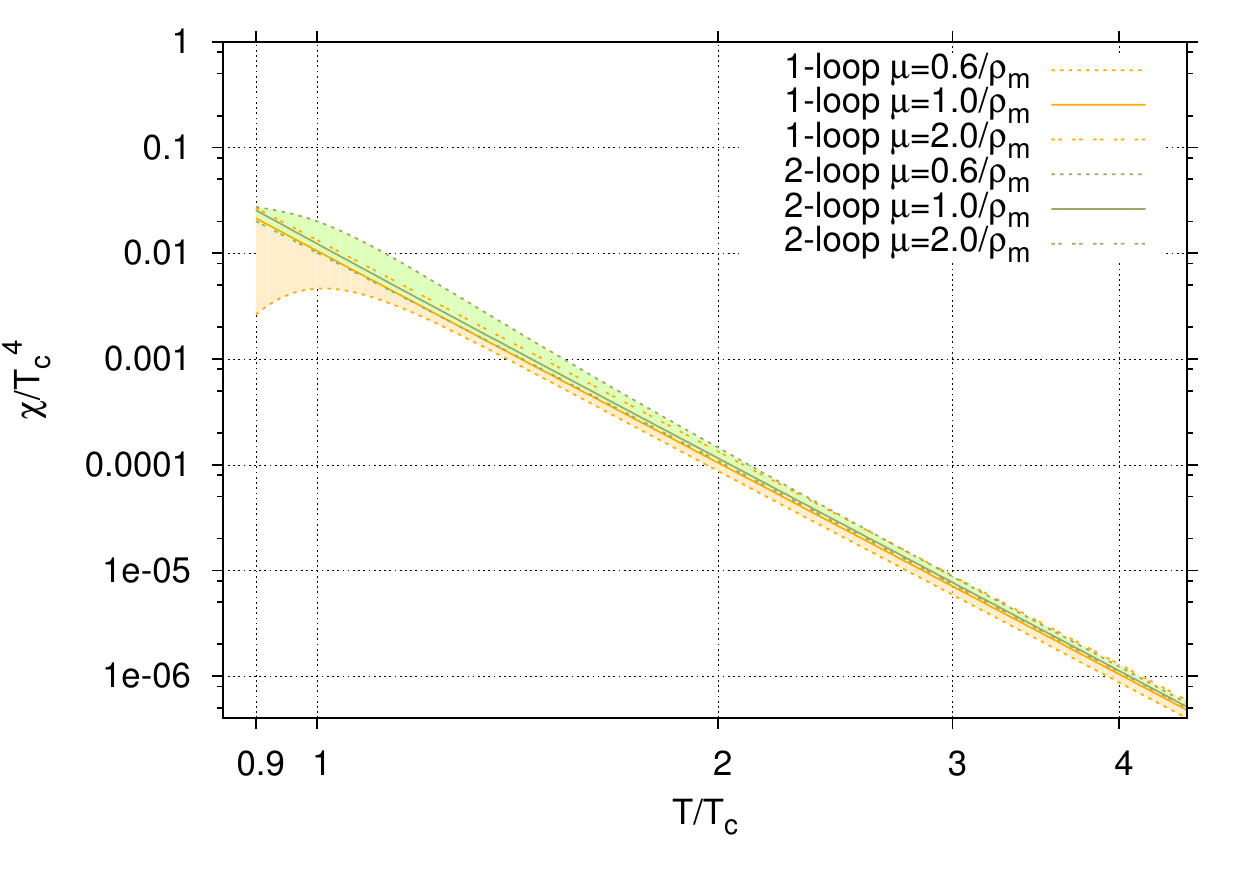}
\caption{Prediction of the topological susceptibility in the 
DIGA: comparison between one-loop and two-loop RGI results.
We used the four-loop expression for the 
running coupling in the modified minimal subtraction scheme as given in the appendix of 
Ref. \cite{Chetyrkin:1997un} and the central value of $T_c/\Lambda^{(n_f=0)}_{\overline{\rm MS}}=1.26(7)$ as determined from
the lattice in Ref. \cite{Borsanyi:2012ve}. 
\label{fi:diga_plot}}
\end{figure}

At finite temperature, electric Debye screening prohibits the existence of large-scale coherent fields in the plasma, leading to the factor~\cite{Pisarski:1980md,Gross:1980br}, 
\begin{equation}
G(x)  \equiv
\exp{\left\{ -2 x^2 
-18 A(x)\right\} },
\end{equation} 
with  
\begin{equation}
A(x)\simeq -\frac{1}{12}\ln \left[ 1+(\pi\rho T)^2/3\right]+\alpha \left[ 1+\gamma(\pi\rho T)^{-3/2}\right]^{-8},
\end{equation}
and $\alpha=0.01289764$ and $\gamma=0.15858$, in Eq.  \eqref{eq:top_susc_diga}. This factor cuts
off the integration over the size distribution in Eq. \eqref{eq:top_susc_diga} at $x = \pi\rho T\sim 1$ and 
ensures the validity of the DIGA at large temperatures, at which $\alpha_s(\pi T)\ll 1$. 

\begin{table}
	\centering
	\begin{tabular}{|c||c|c|c|c|c|}
	\hline
	$T/T_c$ & 1.5 & 2 & 3 & 4 & 5 \\
	\hline
	$b$ ($\kappa = 0.6$) & -6.04 & -6.26 & -6.43 & -6.50 & -6.55 \\
	\hline
	$b$ ($\kappa = 1$) & -6.37 & -6.46 & -6.55 & -6.59 & -6.62 \\
	\hline
	$b$ ($\kappa = 2$) & -6.55 & -6.59 & -6.64 & -6.67 & -6.69 \\
	\hline
	\end{tabular}
	\caption{Temperature slopes of the topological susceptibility predicted in the two-loop RGI DIGA, for a range of renormalization scales according to  Eq.\ \eqref{choice_mu_r}.
\label{ta:diga_slopes}}
\end{table}

\begin{table}
               \centering
               \begin{tabular}{|c||c|c|c|c|c|}
               \hline
               $T/T_c$ & 1 & 2 & 3 & 4 & 5 \\
               \hline
               $\alpha_{\overline{\rm MS}} (1/\rho_m)$ & 0.36 & 0.23 & 0.19 & 0.17 & 0.16 \\
               \hline
               \end{tabular}
               \caption{Strong coupling constant at $\mu_r=1/\rho_m$ for the temperature range covered by the lattice.}
\label{ta:alpha_s}
\end{table}

\begin{figure}[h]
\includegraphics[width=\columnwidth]{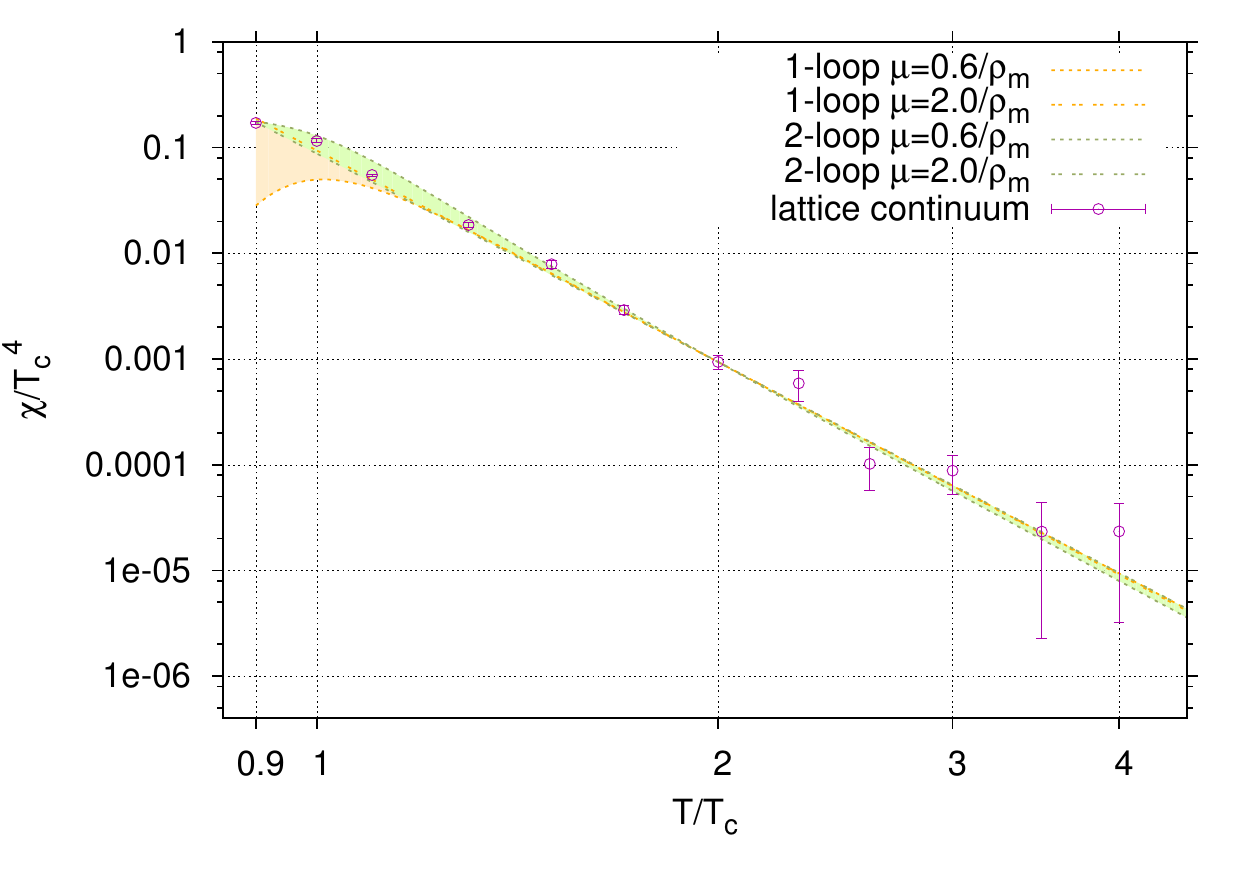}
\caption{Rescaled one-loop and two-loop RGI DIGA results compared to 
lattice continuum extrapolation. The DIGA results shown 
in Fig.\  \ref{fi:diga_plot} were scaled by a factor $K$ of order ten such that they coincide at $T/T_c=2$ with the central value of the lattice continuum data.
\label{fi:comp_plot}}
\end{figure}

Collecting all the factors, the topological susceptibility, in the one-loop DIGA, reads  
\begin{eqnarray}
\label{chi_one_loop}
\chi (T) 
 &\simeq & 2\,
d_{\overline{\rm MS}} \,
 \left( \pi T\right)^{4}
\,
\left( \frac{\mu_r}{\pi T}\right)^{11}   \ I \ 
\left(\frac{2\pi}{\alpha_{\overline{\rm MS}}(\mu_r)}\right)^{6}
 \,  
\\
&&
\times  
\exp{\left(-\frac{2\pi}{\alpha_{\overline{\rm MS}}(\mu_r)}\right)} 
\left[ 1 +\mathcal{O}(\alpha_{\overline{\rm MS}}(\mu_r)) \right]
,
\nonumber
\end{eqnarray}
with 
\begin{eqnarray}
I  
= \int_0^\infty  dx \,x^{6}\,G(x) = 0.267271
.
\end{eqnarray}

This result, however, still suffers from a sizeable dependence on the renormalization scale $\mu_r$, reflecting the importance of the neglected two-loop and higher order contributions. In fact, it is tamed by taking into account the 
ultraviolet part of the two-loop correction. The latter
has been calculated in Ref. \cite{Morris:1984zi} and shown to have exactly the form that 
the gauge coupling becomes a parameter running according to the renormalization group (RG). Therefore, the ultimate, all order result for 
the topological susceptibility becomes independent of $\mu_r$, for $\mu_r \to \infty$. 
At two loop, the corrections amount to a factor 
\begin{eqnarray}
\left(\rho\mu_r\right)^{(\beta_1-12\,\beta_0)\, \alpha_{\overline{\rm MS}}(\mu_r)/(4\pi)}
; \hspace{3ex} \beta_1=102,
\end{eqnarray} 
in $D(\rho )$. Therefore, this RG improvement can be taken into account by 
replacing the factor $I$  in Eq. \eqref{chi_one_loop} by 
\begin{eqnarray}
\tilde I 
&= &
\left( \frac{\mu_r}{\pi T}\right)^{-30\, 
\alpha_{\overline{\rm MS}}(\mu_r)/(4\pi)}
\\ \nonumber
&& \times
\int_0^\infty  dx \,x^{6-30\, 
\alpha_{\overline{\rm MS}}(\mu_r)/(4\pi)}
\,G(x).
\end{eqnarray}
In fact, exploiting the two-loop RG improvement, the $\mu_r$ dependence is heavily reduced, as is obvious
from Fig. \ref{fi:diga_plot}, where we have used as a natural renormalization scale 
\begin{equation}
\mu_r = \kappa/\rho_m = \kappa \pi T/1.2,   
\label{choice_mu_r}
\end{equation}
with $\rho_m$ being approximately the maximum of the integrand of $\tilde I$, and varied the 
remaining free parameter $\kappa$ between 0.6 and 2. 
The renormalization scale dependence appears to be highly reduced
in the regimes $> 3T_c$ and $<T_c$. However, in an 
intermediate region, $\sim T_c$\,--\,$2T_c$, it is comparable in size to the 
one at one-loop.

We present in Table \ref{ta:diga_slopes} the power-law behavior predicted by the two-loop RGI DIGA at various temperatures, which can be compared to the fit \eqref{lattice_fitting_parameters} to the continuum lattice result. 
As far as the overall normalization of the DIGA result for $\chi$ is concerned, one still expects a large uncertainty in the temperature
range available from the lattice. In fact, at these temperatures, $\alpha_s$ is not small, see Tab.\ \ref{ta:alpha_s}. 
Apart from the ultraviolet part, there will be a finite
part of the two-loop correction which will affect mainly the overall normalization of
$\chi$ and will depend on the temperature only logarithmically. Unfortunately, this finite part is not known, yet. Therefore, when comparing to the continuum extrapolated lattice results,
we allow a multiplicative factor $K$ to account for this uncertainty, i.e.\ we absorb the remaining higher loop uncertainties by
replacing
\begin{equation}
\left[ 1 +\mathcal{O}(\alpha_{\overline{\rm MS}}(\mu_r=\kappa\pi T/1.2)) \right] \rightarrow K(T/T_c) 
\end{equation}
in Eq.\ \eqref{chi_one_loop} and the corresponding two-loop RGI expression. Clearly, the $K$-factor should approach unity at very large $T/T_c$.

Figure \ref{fi:comp_plot} nicely illustrates the agreement between the DIGA and the lattice result, if a $K$-factor of order ten is
included\footnote{$K$ factors of order ten to fifty are not uncommon even at next-to-leading order in ordinary perturbative
QCD, see e.g.\ Ref.\ \cite{Rubin:2010xp}.}. More precisely, fitting the lattice continuum data with the rescaled DIGA expression in the temperature range $T/T_c\geq 1$, one finds
\begin{equation}
K = 9.2\pm 0.6\,, \hspace{3ex} {\rm at\ }\ 95\%\,{\rm CL},
\end{equation}
while a fit in the temperature range $T/T_c\geq 2$ yields
\begin{equation}
K = 8.0\pm 3.5\,, \hspace{3ex} {\rm at\ }\ 95\%\,{\rm CL}.
\end{equation}
Apparently, in the temperature range accessible to the lattice, the higher order corrections to the pre-factor of the DIGA are still appreciable, but there are indications of a trend that the 
$K$-factor gets smaller, as 
expected, towards larger values of $T/T_c$.

The $K$-factor strongly depends on the value of $T_c//\Lambda^{(n_f=0)}_{\overline{\rm MS}}$: 
it reduces to one for $T_c//\Lambda^{(n_f=0)}_{\overline{\rm MS}}\simeq 1.03$. However, the latter value is about 3 sigma below 
the central value determined in Ref.\ \cite{Borsanyi:2012ve}, $T_c//\Lambda^{(n_f=0)}_{\overline{\rm MS}}\simeq 1.26(7)$.

\section{Conclusions}

This paper presents lattice and DIGA calculations of the $\theta$ and 
temperature dependence of the free energy density  of QCD in the quenched 
limit, i.e.\ for infinite quark masses. In the lattice approach the 
temperature ranges from $0.9 T_c$ to $4 T_c$ and thus significantly 
extends former results. 
The precise 
high temperature data allows to compare the lattice results with first 
principle calculations within the DIGA. For the evaluation of $\chi (T)$ 
we use the two-loop RGI version of the instanton size distribution, which 
is less sensitive with respect to the renormalization scale. We find that 
the two different methods show nice agreement after one includes an overall 
correction factor in the DIGA results. This order ten correction is supposed 
to take into account higher loop contributions in the normalization, similar 
to so-called $K$-factors in perturbative QCD processes. These are significant,
since in the considered temperature region the strong coupling 
constant is still large. Nevertheless, the achieved results 
motivate an investigation of more realistic models involving light quarks.

\begin{figure}
\begin{minipage}{\columnwidth}
\centering
\includegraphics[width=0.94\columnwidth]{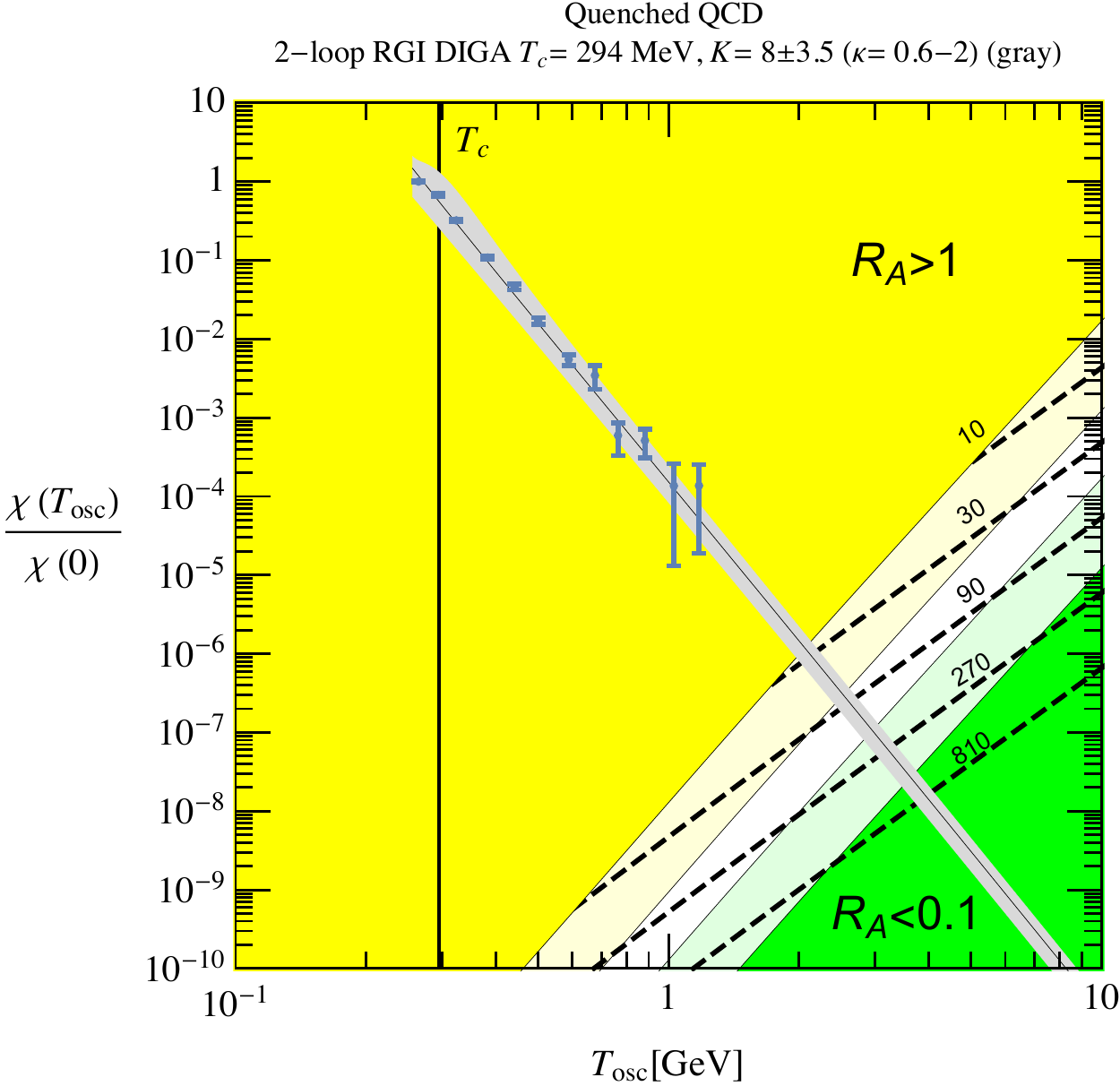}\\[4ex]
\includegraphics[width=0.94\columnwidth]{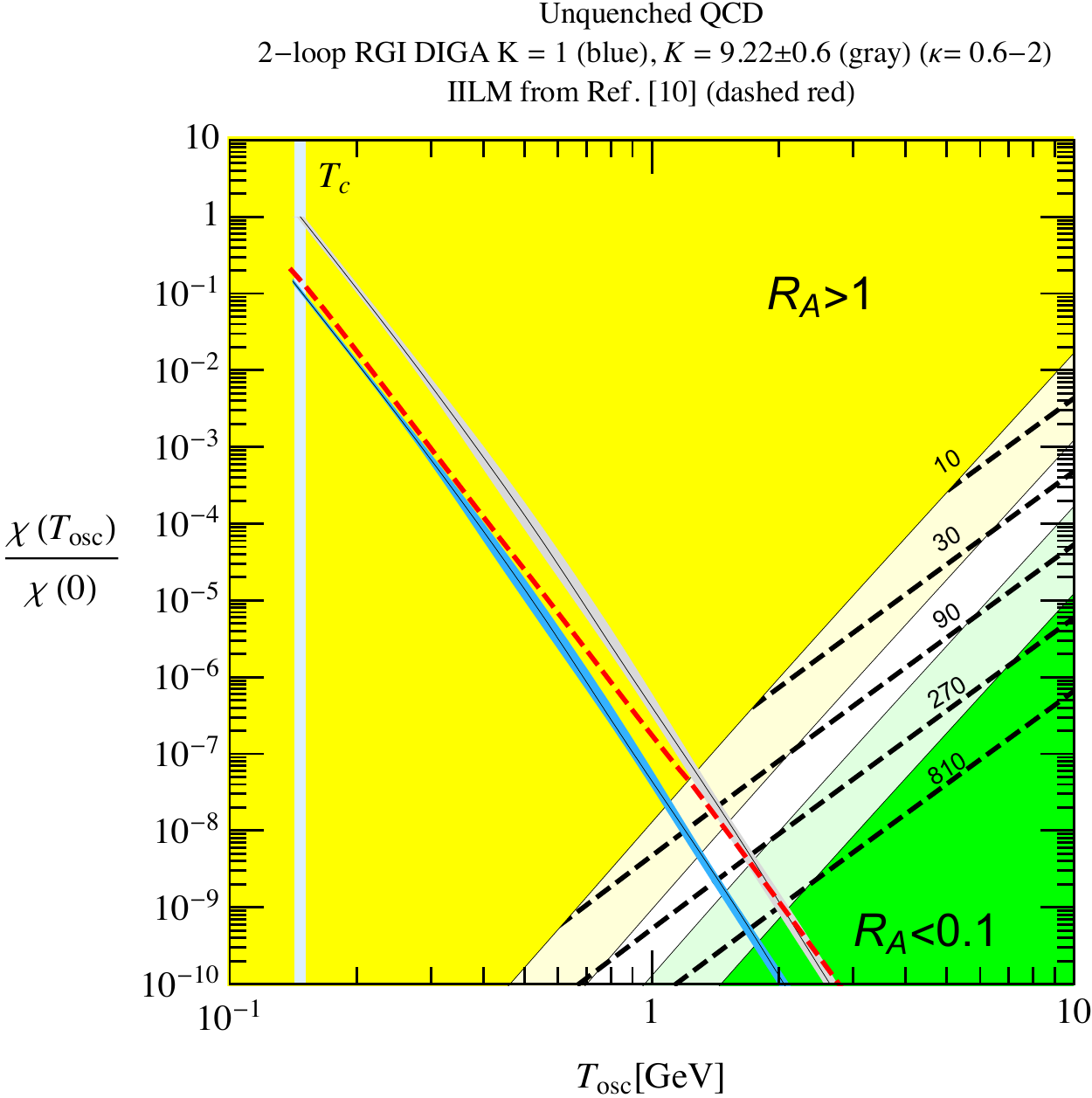}
\end{minipage}
\caption{Consequence of our findings for axion dark matter 
from quenched (top) and full QCD (bottom). The dark yellow
region is excluded because $R_A>1$ even without a contribution from
strings. The light yellow region indicates
$R_A>1$ when string contributions are included. 
In the dark green region axions even including strings give a
small contribution ($R_A<0.1$) to dark matter. 
The light green region indicates $R_A<0.1$ just from the
misalignment mechanism. In order to compare the quenched results with the 
axion dark matter scenario we had to transform dimensionless quantities 
into dimensionful ones. This is not unambiguous since we do not have pure 
gauge theory in nature. Using different quantities to set the scale one 
gets $280$ -- $310$~MeV for $T_c$. For illustrative purposes 
we use $T_c=294$~MeV which lies in the middle of the range and is a 
factor two larger than the full QCD transition temperature. Our lattice results
are shown by the blue points and the two-loop RGI DIGA prediction by the gray band in the quenched Figure.
In the unquenched case the blue and gray bands correspond to the two-loop RGI
DIGA predictions with $K=1$ and $K=9.22 \pm 0.6$, respectively. The dashed red line
shows the IILM prediction of Refs.~\cite{Wantz:2009it,Wantz:2009mi}.
\label{fi:chi_vs_T_R}}
\end{figure}

The possible consequences of our findings on the predictions of the amount of axion dark matter can be read off from 
Fig.\ \ref{fi:chi_vs_T_R}. 
First, we have plotted the regions where axion DM would be overproduced with respect to observations and are thus excluded. The darker yellow region is excluded just by the axions produced from the misalignment mechanism, 
cf. Eq.\ \eqref{misalignment}, assuming a flat distribution of initial misalignment angles in the observable universe\footnote{In the scenario where inflation homogenizes one particular value of $\theta$ across our universe, a prediction of the axion DM abundance is not possible, except perhaps in an anthropic sense~\cite{Preskill:1982cy,Pi:1984pv} } $\theta\in [-\pi,\pi]$. For the DIGA potential, $\chi(T)(1-\cos(\theta))$, we calculate $\langle\theta^2\rangle f(\langle\theta^2\rangle)\simeq 6.6$ from numerically evolving the axion field with different initial conditions. The exclusion extends to the light yellow region if we consider the DM axions produced from the decay of unstable axionic strings and domain walls according to Ref. \cite{Kawasaki:2014sqa}. 
The excluded region has to be compared with our lattice results (blue dots) and the two-loop DIGA calculation with the fitted $K$-factor, for which we have set the scale $T_c=294$ MeV (twice the dynamical value) for illustration purposes, see  Fig.\ \ref{fi:chi_vs_T_R} (top) . Despite the arbitrariness we can derive a number of interesting conclusions. First note that our results correspond to a region where axion DM does not offer a viable cosmology because $R_A>1$. However, our lattice results are too short in $T/T_c$ only by a factor of 2--3. Further developments from the lattice side could in principle lead to a direct estimation of the axion mass. Meanwhile, the DIGA result allows a 
controlled extrapolation of our results over the interesting region, 
that can be used to set a constraint (here merely {\em illustrative}): 
$25\; \mu\rm eV \lesssim m_A \lesssim 525\; \mu \rm eV$ assuming that
at least ten percent of dark matter can be attributed to axions. 

For the final calculation of the QCD axion abundance as dark matter candidate 
one needs to include the light quarks. This complicates the lattice evaluation
by far. However, the DIGA approximation can still be carried out and 
self-consistently takes into account the running of the quark masses as 
well as their influence on the strong coupling constant. We present this 
preliminary result as a blue band in Fig.\ \ref{fi:chi_vs_T_R} (bottom). 
From what we learned in the quenched case, this DIGA result still misses an 
O(10) constant factor but the lack of lattice data in this case prevents 
from fitting the expected $K-$factor explicitly. 
We can tentatively estimate it to be the required factor that 
makes $\chi(T)/\chi(0)=1$ at $T_c$. Since the transition in full QCD is a crossover~\cite{Aoki:2006we},
$T_c$ is not unambiguous. Using the transition temperature defined by the
chiral susceptibility, $T_c=147(2)(3)$~MeV~\cite{Aoki:2006br,Aoki:2009sc,Borsanyi:2010bp}, 
we obtain $K=9.22\pm 0.6$. Very interestingly, this agrees with the IILM model calculation of Ref. \cite{Wantz:2009it,Wantz:2009mi} (red-dashed line) at the temperatures of interest. 
This factor of $9.22$ and the previous $8$ of the quenched limits do not 
affect very strongly the extraction of values of the axion mass, owing to 
the strong $T$-dependence of $\chi(T)$. 
Using $K=9.22$ and again assuming at least ten percent axion
contribution to dark matter we get the 
range $40\;\mu\rm eV\lesssim m_A\lesssim 930\;\mu\rm eV$
while using $K=1$ we would get $50\;\mu\rm eV\lesssim m_A\lesssim 1100\;\mu\rm eV$, 
i.e. only a $\sim 20\%$ correction for an ${\cal O}(10)$ $K$ factor uncertainty.

\section*{Acknowledgments}

The authors wish to thank G. Moore for useful discussions.
Computations were performed on JUQUEEN at FZ-J\"ulich and on GPU clusters
at Wuppertal and Budapest. This project was funded by the DFG grant SFB/TR55 and by OTKA under grant
OTKA-NF-104034. S.D.K. is funded
by the "Lend\"ulet" program of the Hungarian Academy of Sciences (LP2012-44/2012).
M.D. acknowledges support from the Studienstiftung des Deutschen Volkes and J. R. from 
the  Ramon y Cajal Fellowship 2012-10597 from the  Spanish Ministry of Economy and Competitivity. 

\section*{References}

\bibliography{paper1}

\end{document}